\documentclass[twocolumn]{bmcart}% uncomment this for twocolumn layout and comment line below
\usepackage{amsthm,amsmath,amssymb}
\usepackage[utf8]{inputenc} %unicode support

\usepackage{graphicx}
\usepackage{booktabs}
\usepackage{colortbl}
\usepackage{hyperref}
\usepackage{cite}

\startlocaldefs

\graphicspath{ {./figures/} }

\def\perec{p(E_{rec})} %fraction of reciprocal event pairs
\def\plrec{p(l_{rec})} %fraction of reciprocal links

\newcommand{\fref}[1]{Fig.~\ref{#1}}

\newcommand{\tref}[1]{Table~\ref{#1}}

\newcommand{\srefsi}[1]{SI Section~S{#1}}

\endlocaldefs

\begin{document}

%%% Start of article front matter
\begin{frontmatter}

\begin{fmbox}
\dochead{Research}


\title{Temporal patterns of reciprocity in communication networks}

%%%%%%%%%%%%%%%%%%%%%%%%%%%%%%%%%%%%%%%%%%%%%%
%%                                          %%
%% Enter the authors here                   %%
%%                                          %%
%% Specify information, if available,       %%
%% in the form:                             %%
%%   <key>={<id1>,<id2>}                    %%
%%   <key>=                                 %%
%% Comment or delete the keys which are     %%
%% not used. Repeat \author command as much %%
%% as required.                             %%
%%                                          %%
%%%%%%%%%%%%%%%%%%%%%%%%%%%%%%%%%%%%%%%%%%%%%%

\author[
   addressref={ceu},
   noteref={n1}, 
]{\inits{SC}\fnm{Sandeep} \snm{Chowdhary}
}
\author[
   addressref={ceu},
   noteref={n1}, 
]{\inits{EA}\fnm{Elsa} \snm{Andres}
}
\author[
   addressref={ceu},
   noteref={n1}, 
]{\inits{AM}\fnm{Adriana} \snm{Manna}
}
\author[
   addressref={ceu},
   noteref={n1}, 
]{\inits{LB}\fnm{Luka} \snm{Blagojevi\'c}
}
\author[
   addressref={ceu},
]{\inits{LDG}\fnm{Leonardo} \snm{Di Gaetano}
}
\author[
   addressref={ceu,aalto,unam}, 
   corref={ceu}, 
   email={iniguezg@ceu.edu}
]{\inits{GI}\fnm{Gerardo} \snm{Iñiguez}}

%%%%%%%%%%%%%%%%%%%%%%%%%%%%%%%%%%%%%%%%%%%%%%
%%                                          %%
%% Enter the authors' addresses here        %%
%%                                          %%
%% Repeat \address commands as much as      %%
%% required.                                %%
%%                                          %%
%%%%%%%%%%%%%%%%%%%%%%%%%%%%%%%%%%%%%%%%%%%%%%

\address[id=ceu]{
  \orgname{Department of Network and Data Science, Central European University}, % university, etc
  \postcode{1100}
  \city{Vienna}, 
  \cny{Austria}
}
\address[id=aalto]{
  \orgname{Department of Computer Science, Aalto University School of Science},
  \postcode{00076}
  \city{Aalto},
  \cny{Finland}
}
\address[id=unam]{
  \orgname{Centro de Ciencias de la Complejidad, Universidad Nacional Autonóma de México},
  \postcode{04510}
  \city{Ciudad de México},
  \cny{Mexico}
}

%%%%%%%%%%%%%%%%%%%%%%%%%%%%%%%%%%%%%%%%%%%%%%
%%                                          %%
%% Enter short notes here                   %%
%%                                          %%
%% Short notes will be after addresses      %%
%% on first page.                           %%
%%                                          %%
%%%%%%%%%%%%%%%%%%%%%%%%%%%%%%%%%%%%%%%%%%%%%%

\begin{artnotes}
\note[id=n1]{Equal contributor}
\end{artnotes}

%\end{fmbox}% comment this for two column layout

%%%%%%%%%%%%%%%%%%%%%%%%%%%%%%%%%%%%%%%%%%%%%%
%%                                          %%
%% The Abstract begins here                 %%
%%                                          %%
%% Please refer to the Instructions for     %%
%% authors on http://www.biomedcentral.com  %%
%% and include the section headings         %%
%% accordingly for your article type.       %%
%%                                          %%
%%%%%%%%%%%%%%%%%%%%%%%%%%%%%%%%%%%%%%%%%%%%%%

\begin{abstractbox}

\begin{abstract} % abstract
Human communication, the essence of collective social phenomena ranging from small-scale organizations to worldwide online platforms, features intense reciprocal interactions between members in order to achieve stability, cohesion, and cooperation in social networks. While high levels of reciprocity are well known in aggregated communication data, temporal patterns of reciprocal information exchange have received far less attention. Here we propose measures of reciprocity based on the time ordering of interactions and explore them in data from multiple communication channels, including calls, messaging and social media. By separating each channel into reciprocal and non-reciprocal temporal networks, we find persistent trends that point to the distinct roles of one-on-one exchange versus information broadcast. We implement several null models of communication activity, which identify memory, a higher tendency to repeat interactions with past contacts, as a key source of reciprocity. When adding memory to a model of activity-driven, time-varying networks, we reproduce the levels of reciprocity seen in empirical data. Our work adds to the theoretical understanding of the emergence of reciprocity in human communication systems, hinting at the mechanisms behind the formation of norms in social exchange and large-scale cooperation.
\end{abstract}

%%%%%%%%%%%%%%%%%%%%%%%%%%%%%%%%%%%%%%%%%%%%%%
%%                                          %%
%% The keywords begin here                  %%
%%                                          %%
%% Put each keyword in separate \kwd{}.     %%
%%                                          %%
%%%%%%%%%%%%%%%%%%%%%%%%%%%%%%%%%%%%%%%%%%%%%%

\begin{keyword}
\kwd{Reciprocity}
\kwd{Temporal networks}
\kwd{Human communication}
\end{keyword}

% MSC classifications codes, if any
%\begin{keyword}[class=AMS]
%\kwd[Primary ]{}
%\kwd{}
%\kwd[; secondary ]{}
%\end{keyword}

\end{abstractbox}
\end{fmbox}% uncomment this for twcolumn layout

\end{frontmatter}

%%%%%%%%%%%%%%%%%%%%%%%%%%%%%%%%%%%%%%%%%%%%%%
%%                                          %%
%% The Main Body begins here                %%
%%                                          %%
%% Please refer to the instructions for     %%
%% authors on:                              %%
%% http://www.biomedcentral.com/info/authors%%
%% and include the section headings         %%
%% accordingly for your article type.       %%
%%                                          %%
%% See the Results and Discussion section   %%
%% for details on how to create sub-sections%%
%%                                          %%
%% use \cite{...} to cite references        %%
%%  \cite{koon} and                         %%
%%  \cite{oreg,khar,zvai,xjon,schn,pond}    %%
%%  \nocite{smith,marg,hunn,advi,koha,mouse}%%
%%                                          %%
%%%%%%%%%%%%%%%%%%%%%%%%%%%%%%%%%%%%%%%%%%%%%%

%%%%%%%%%%%%%%%%%%%%%%%%% start of article main body
% <put your article body there>

\section*{Introduction}

Reciprocity,  the tendency of entities to mutually interact,  is a widespread feature of complex networked systems,  central to social network analysis \cite{wasserman1994social,mandel2000measuring},  evolutionary game theory \cite{trivers1971evolution,binmore1994game,nowak2005evolution},  and the economics of public goods and social norms \cite{berg1995trust,fehr2000fairness}.  Already recognized in some of the earliest sociometrics studies \cite{moreno1938statistics},  reciprocity is an emergent moral norm of human interaction \cite{gouldner1960norm} indicating stability,  cohesion,  and cooperation in social networks \cite{simmel1950sociology,friedkin2004social,nowak2006five,molm2007value},  which contributes to tie strength \cite{granovetter1973strength} and social influence \cite{mahmoodi2018reciprocity}.

Highly reciprocal patterns of connectivity have been found in static,  aggregated data from the world trade web,  internet, and neurons \cite{garlaschelli2004patterns},  and in social networks of communication \cite{kovanen2010reciprocity,akoglu2012quantifying,karsai2012correlated},  kinship \cite{schnegg2006reciprocity},  and strategic partnerships \cite{wincent2010quality}. This has prompted the development of reference models with tunable amounts of reciprocity within the framework of exponential random graphs \cite{holland1981exponential,park2004statistical,snijders2011statistical},  both in the absence \cite{garlaschelli2006multispecies} and presence \cite{zamora2008reciprocity,zlatic2009influence} of degree correlations.  The resulting reciprocity measures have been extended to weighted \cite{fagiolo2006directed,kovanen2010reciprocity,wang2013dyadic,squartini2013reciprocity} and bipartite \cite{zhao2013user} networks,  and used to uncover the role of reciprocal links in the world wide web \cite{serrano2007decoding,perra2009pagerank}, the growth of Wikipedia \cite{zlatic2006wikipedias,zlatic2011model}, synchronization in brain networks \cite{zhou2006hierarchical},  and the dynamics of scientific reputation \cite{li2019reciprocity}.

When inferring social network structure from repeated interactions like communication events \cite{onnela2007structure,urena2020estimating},  however,  reciprocity emerges as an inherently temporal property.  A scenario in which individual $A$ receives 10 messages from person $B$, followed by 10 messages from $B$ to $A$,  is structurally different from the case where 20 messages are exchanged in an alternating way ($A \to B$, $B \to A$, etc.).  An appropriate framework is that of temporal networks \cite{holme2012temporal,holme2015modern}, where nodes are people and time-stamped edges are events in potentially multiple communication channels (face-to-face,  calls,  text,  email, online messaging,  social media, etc.) \cite{vlahovic2012effects,wang2013usage,quadri2014multidimensional}.  In contrast to the static case,  reciprocity in temporal, non-aggregated network data has received less attention in the literature.  Notable exceptions are the extension of reciprocity measures to spatio-temporal urban networks \cite{williams2016spatio},  as well as studies of the role of reciprocity in the temporal stability of non-human social networks \cite{dakin2020reciprocity}, and in the dynamics of both collaboration \cite{brandenberger2018trading} and organizational \cite{quintane2013short,kitts2017investigating} networks.

Here we explore the temporal patterns of reciprocity in social networks by analyzing communication data in several channels \cite{stopczynski2014measuring,sapiezynski2019interaction,panzarasa2009patterns,paranjape2017motifs}. We start by proposing measures of reciprocity that explicitly take into account the time ordering of events and are thus related to widely studied patterns of temporal inhomogeneity like burstiness \cite{barabasi2005origin,unicomb2021dynamics}.  These measures give additional information than their aggregated counterparts \cite{garlaschelli2004patterns,squartini2013reciprocity}, particularly the overall balance between events in different directions over a social tie \cite{karsai2012correlated}.  By separating each dataset into reciprocal and non-reciprocal temporal networks, we observe persistent differences between channels that point to their distinct roles in communication \cite{jensen2011internet}, in agreement with previous work on the structure of egocentric networks \cite{saramaki2014persistence,heydari2018multichannel} and daily patterns of communication \cite{aledavood2016channel}.  Finally,  we introduce a model within the framework of activity-driven,  time-varying networks \cite{perra2012activity,liu2014controlling}, combining both heterogeneous node activity \cite{pozzana2017epidemic} and repeated interactions over established social connections \cite{karsai2014time,kim2018dynamic}, which recovers the empirical levels of reciprocity seen in temporal communication networks.

\begin{figure*}[t]
\includegraphics[scale=.32,trim={0 12.1cm 12cm 1.5cm},clip]{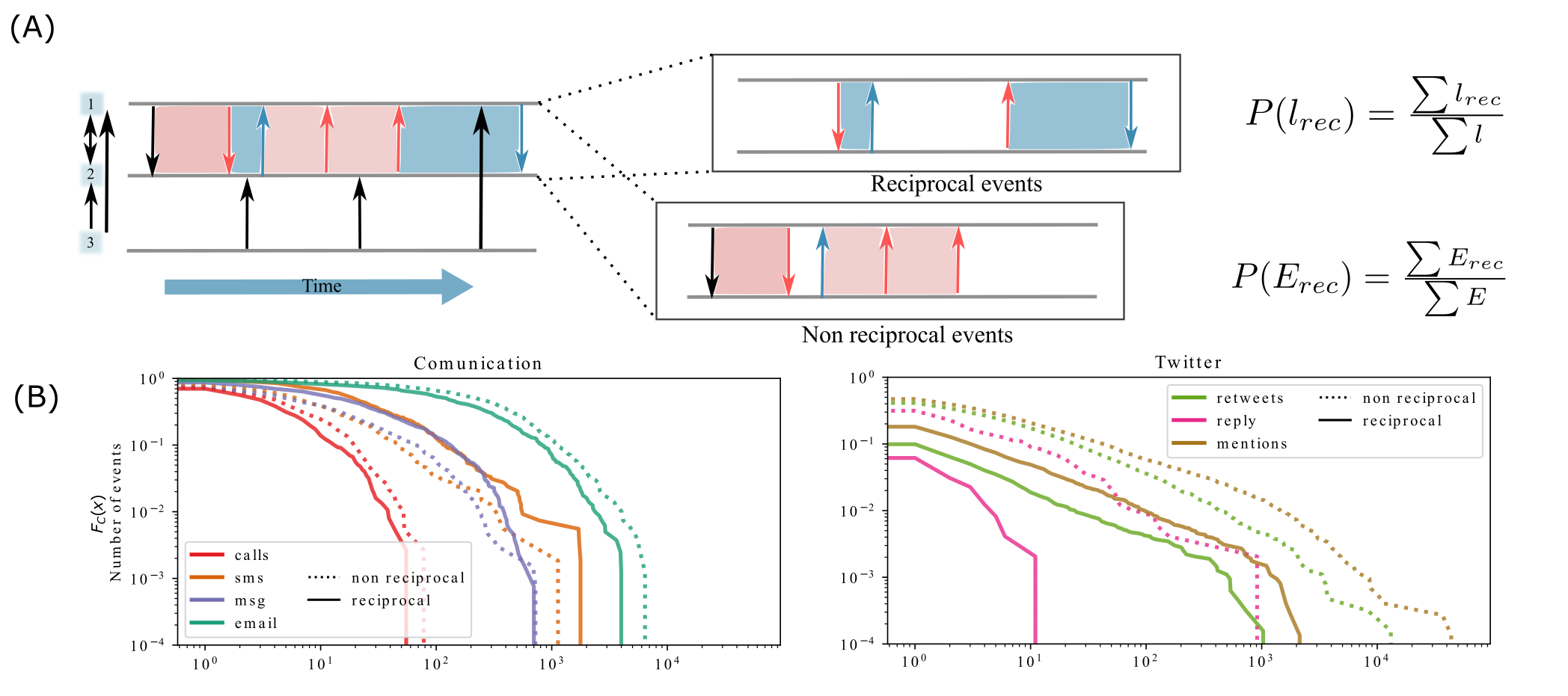}
\includegraphics[scale=.83,trim={0 0.3cm 0 0cm},clip]{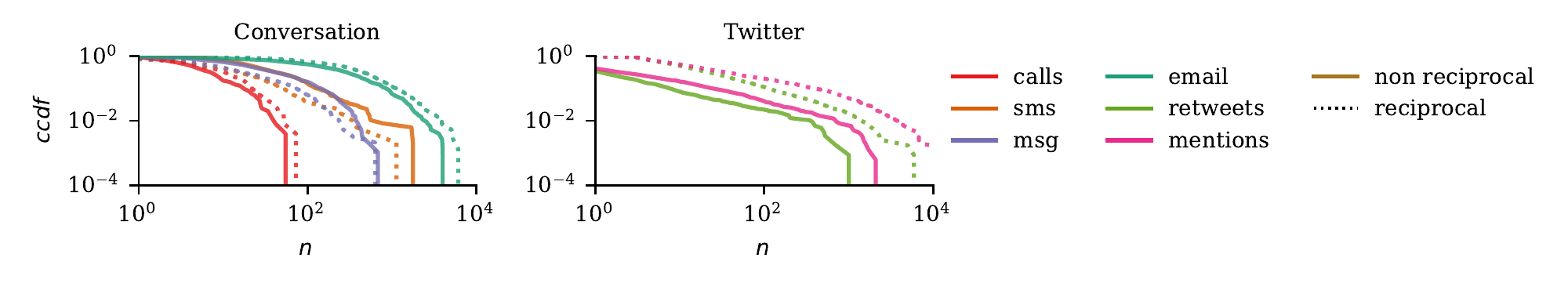}
\caption{\textbf{Reciprocal and non-reciprocal activity in empirical communication networks.} {\bf (top)} Schema for separating a temporal network of time-stamped, directed communication events between pairs of individuals into reciprocal and non-reciprocal components. A pair of consecutive events is a {\it reciprocation} if events have the opposite direction (blue areas), and a {\it non-reciprocation} if they have the same direction (red areas). We measure the fraction of reciprocations over a link, $\perec$, and the fraction of links having at least one reciprocation, $\plrec$ (see \srefsi{1}). {\bf (bottom)} Complementary cumulative distribution function $P(n' > n)$ (ccdf), the fraction of nodes having strictly more than $n$ reciprocations (solid) or non-reciprocations (dashed) in various communication channels. One-on-one communication channels tend to be more reciprocal than broadcasting channels (i.e. Twitter; see \tref{tab_}).}
\label{fig:fig1}
\end{figure*}
%\san{Mark Time-gaps and \emph{Time-gap burstiness} in top panel.}

\section*{Results}

\begin{table}
\begin{tabular}{{|l|l|l|l|l|l|l|}}
\hline
\toprule
\rowcolor[HTML]{EEEEEE} 
\multicolumn{1}{|l|}{\textbf{dataset}}&
\multicolumn{1}{|l|}{$E$} & \multicolumn{1}{l|}{$L$} & 
\multicolumn{1}{l|}{$N$} & \multicolumn{1}{l|}{$\perec$} & \multicolumn{1}{l|}{$\plrec$} \\ 
\rowcolor[HTML]{FFFFFF}
calls & 2430  & 181  & 252  & 0.44  & 0.95  \\
\rowcolor[HTML]{FFFFFF}
sms & 23779  & 473  & 482  & 0.74  & 0.99  \\
\rowcolor[HTML]{FFFFFF}
msg & 40600  & 3343  & 941  & 0.67  & 0.87  \\
\rowcolor[HTML]{FFFFFF}
email & 306529  & 6864  & 753  & 0.45  & 0.90 \\
\rowcolor[HTML]{FFFFFF}
retweets &  57899  & 3142  & 1156  & 0.10  & 0.33 \\
\rowcolor[HTML]{FFFFFF}
mentions & 226774  & 8292  & 1609  & 0.11  & 0.40 \\ 
\bottomrule
\hline
\end{tabular}
\caption{\textbf{Basic statistics of studied datasets}. Temporal network data on calls and messages from the Copenhagen Network Study \cite{stopczynski2014measuring,sapiezynski2019interaction} (calls \& sms), online social network messages at the University of California, Irvine \cite{panzarasa2009patterns} (msg), emails at a European research institution \cite{paranjape2017motifs} (email), and our crawl of keyword-restricted retweets and mentions in Twitter (retweets \& mentions) (see \srefsi{2}) . Table shows the number of events $E$, links $L$, and nodes $N$, the probabilities of having a reciprocal link, $\plrec$, and a reciprocation, $\perec$, and standard burstiness $B$ \cite{goh2008burstiness} (see \srefsi{1}). Most channels (apart from Twitter) show significant levels of reciprocity.
}
\label{tab_}
\end{table}

\begin{figure}
\centering
\includegraphics[scale=0.65]{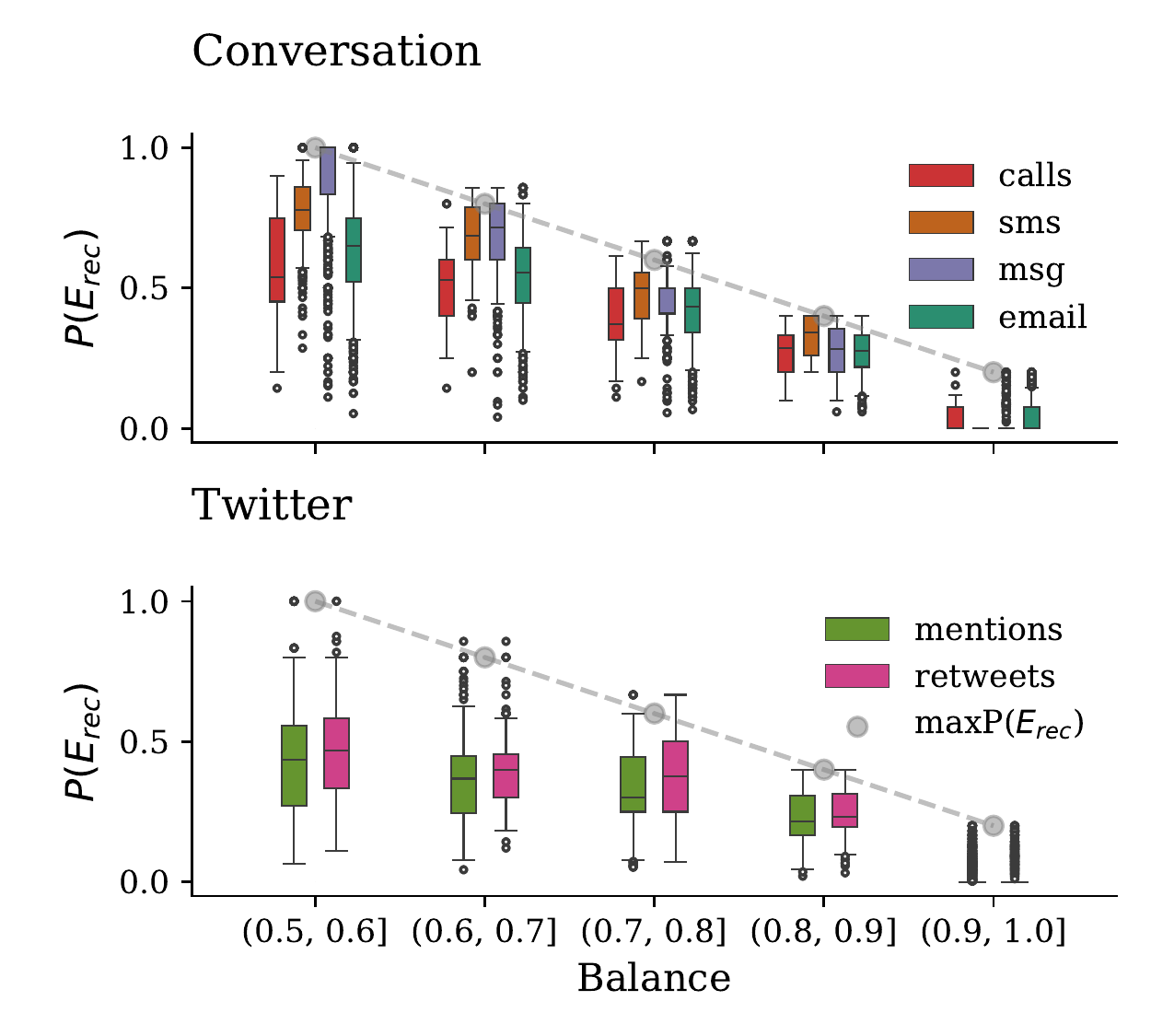}
\caption{\textbf{Balance as upper bound of reciprocity in temporal networks.}  Distribution of fraction of reciprocations $\perec$ over nodes in dataset (box plot), and approximate upper bound $\perec = 2(1-b)$ (see \srefsi{7}), both as function of balance $b$ over nodes \cite{karsai2012correlated}. Balance is constrained to $0.5 \leq b \leq 1$, with $b=1$ an unbalanced, unidirectional relationship from one node to the other, and $b=0.5$ perfect bidirectionality between two nodes. $\perec$ decreases as communication moves away from perfect balance, for both conversation (calls, sms, msg, email) and broadcasting (Twitter) channels. $\perec$ is highest for balanced conversations ($b \in (0.5,0.6]$), and smallest for severely unbalanced interactions between node pairs ($b \in (0.9,1]$). As an aggregate measure, balance does not convey the same information as reciprocity, since $\perec$ varies greatly among the 6 datasets for the same interval of $b$; balance is a necessary but not sufficient condition for reciprocation.  Outliers above the upper bound of balance are due to an approximation in its derivation (see \srefsi{7}).}
\label{fig:fig2}
\end{figure}

\subsection*{Multi-channel communication networks are reciprocal}

We study temporal network data in several communication channels: phone-enabled social interactions via calls and messages in the Copenhagen Network Study \cite{stopczynski2014measuring,sapiezynski2019interaction} (denoted calls \& sms), private messages sent in an online social network at the University of California, Irvine \cite{panzarasa2009patterns} (msg), emails exchanged among members of a European research institution \cite{paranjape2017motifs} (email), and our own crawl of retweets and mentions in Twitter with keywords associated to the anti-vaccination movement in Italy (retweets \& mentions) (\fref{fig:fig1} and \tref{tab_}; for data description see Supplementary Information [SI] Section 2).

In a temporal network of social interactions via communication, two individuals $i$ and $j$, or nodes, interact through a directed time-stamped event $e_{ijt}$, when source node $i$ communicates with target node $j$ at time $t$ (e.g., calls, sends a message, etc.). The time-ordered sequence of events of link $l_{ij}$ is, e.g.,  $\{e_{ijt_1},e_{jit_2},e_{ijt_3}...e_{jit_T}\}$ (with $T$ the total number of events in the link) and one can display its directed events by arrows (\fref{fig:fig1} top; for definitions see \srefsi{1}). Communication between a pair of individuals can then be divided into reciprocal and non-reciprocal components. Two consecutive events in opposite directions form a {\it reciprocation} [($e_{ijt_1}, e_{jit_2}$) with $t_2 > t_1$ ], while two in the same direction are a {\it non-reciprocation} [($e_{ijt_1}, e_{ijt_2}$) with $t_2 > t_1$ ].

We compute the complementary cumulative distribution function $P(n' > n)$ (ccdf), i.e. the fraction of nodes having strictly more than $n$ reciprocations or non-reciprocations in each of the 6 studied communication channels (\fref{fig:fig1} bottom). At this level of aggregation, calls and email are slightly more reciprocal, while sms and msg tend towards non-reciprocity. In both retweets and mentions, Twitter is markedly more non-reciprocal than other communication networks. This contrast is likely due to the different purposes for which these social networks are used \cite{jensen2011internet}. Communication networks (msg, calls, email, sms) are primarily conversation channels where interactions are parts of a discussion, people reaching out to each other and responding throughout time. On the other hand, Twitter is mostly used as a broadcasting platform, where users post to reach the community and do not target specific users.

We begin to explore the temporal nature of reciprocity by measuring the number of reciprocations $E_{rec,ij}$ over link $l_{ij}$, relative to the number of consecutive event pairs on that link, $E_{ij}-1$. By averaging over links, we obtain the reciprocation probability $\perec = \langle E_{rec,ij} / (E_{ij}-1)\rangle_{ij}$. We also compute the number of links with at least one reciprocation ($l_{rec}$) relative to the total number of links ($L$), $\plrec = l_{rec} / L$ (\tref{tab_}). We filter out links with less than five events, the lowest threshold value that starts showing relatively low variation in most quantities studied (for sensitivity analysis see \srefsi{3}). This choice of filtering is motivated by previous studies on social network structure \cite{granovetter1973strength,marsden1984measuring,wang2013dyadic}, which show that repeated interaction is a good proxy for tie strength. In our case, we remove the weakest ties to focus on more persistent patterns of communication.

All conversation channels (calls, sms, msg, email) show high levels of reciprocity. The fraction of reciprocations $\perec$ ranges between 0.74 (sms) and 0.44 (calls, email) (\tref{tab_}). In contrast, low levels of reciprocity in Twitter are likely due to the broadcasting, uni-directional nature of the platform, with $\perec \sim 0.10$. The aggregated network of Twitter shows a significant negative correlation between in- and out-degrees (see \srefsi{5}), meaning that communication between pairs of nodes is potentially unbalanced on aggregate. As we describe in more detail below, if communication between two nodes is highly skewed in one direction, then reciprocity [as measured by $\perec$] cannot be high. We observe a similar behaviour with $\plrec$: most of the links (87--99\%) in conversation channels have at least one reciprocation, while this is only the case for 33--40\% of the links in retweets and mentions.

A way of highlighting the temporal nature of reciprocity is by comparing it with the overall balance between events in different directions over a social tie. Following \cite{karsai2012correlated}, we define balance between nodes $i$ and $j$ as $b_{ij} = \frac{\text{max}(n_{ij}, n_{ji})}{n_{ij} + n_{ji}}$,
where $n_{ij}$ and $n_{ji}$ are the number of events from $i$ to $j$ and from $j$ to $i$, respectively, for link $l_{ij}$ in the aggregated network. In other words, balance quantifies how much the interaction between two individuals is skewed in one direction or another.

\begin{figure*}[t]
\centering
\includegraphics[scale=.7]{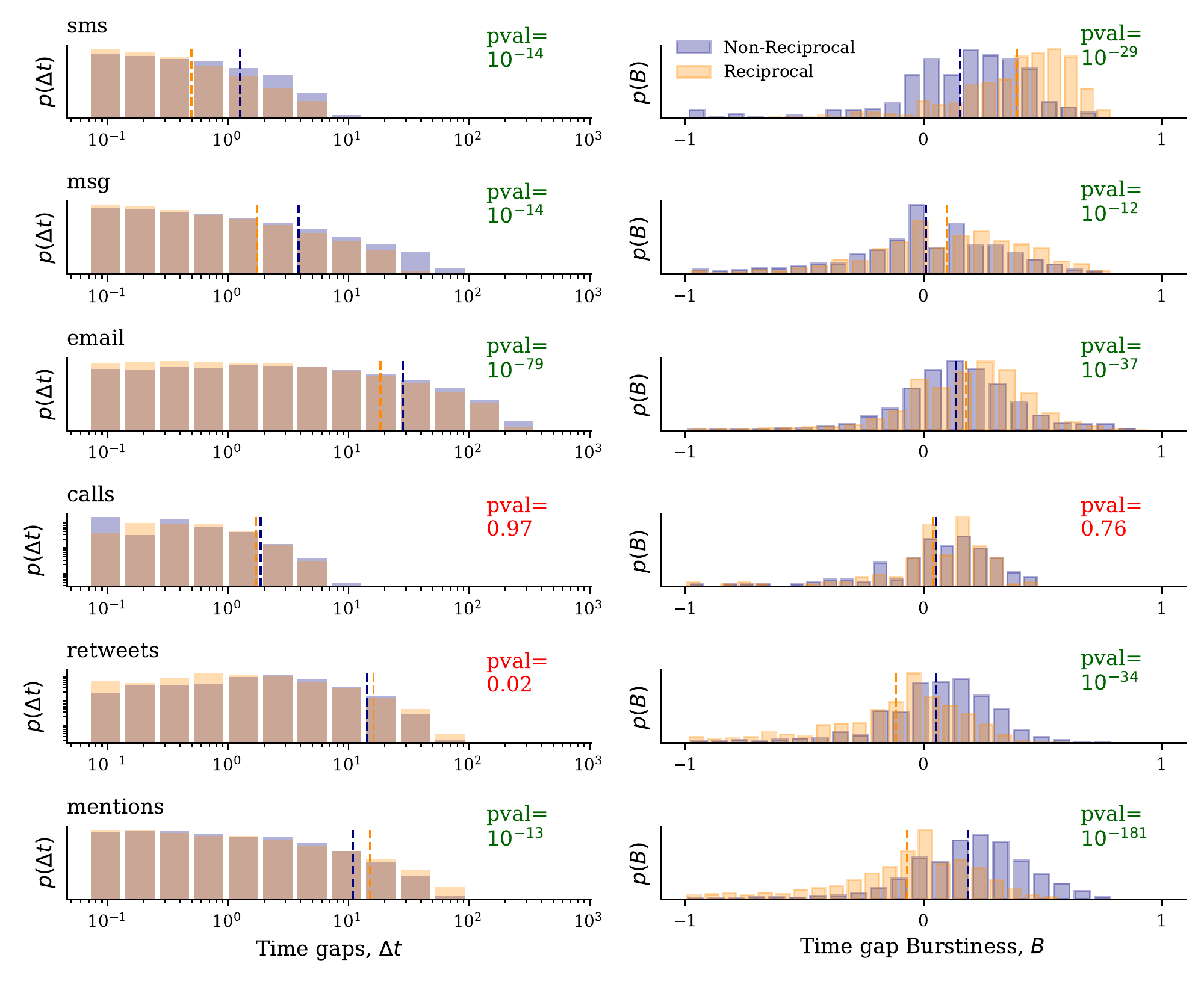}
\caption{\textbf{Reciprocation is more bursty than non-reciprocation in human communication, and varies across channels.} Distribution $p(\Delta t)$ of the time gap  $\Delta t$ (in days) between successive (non-)reciprocations (left column), and distribution $p(B)$ of the time gap burstiness $B$ (right column), both over links of all communication channels. Mean values $\langle \Delta t \rangle$ and $\langle B \rangle$ are marked by dashed lines (same colors as corresponding histograms). Time gaps between successive reciprocations are smaller than between non-reciprocations for conversation channels (calls, sms, msg, emails), while the opposite holds for Twitter.  Reciprocal communication is significantly more bursty (non-homogeneous) than non-reciprocity for sms, msg and email, and becomes less so for calls and the broadcasting channels retweets and mentions. We compute statistical significance of the difference between two distributions via a Kolmogorov-Smirnov (KS) 2-sample test; p-values (pval) $<0.01$ are deemed significant (green, otherwise red).}
\label{fig:fig3}
\end{figure*}

Communication data shows an inverse correlation between balance $b$ in the aggregated network and the fraction of reciprocations $\perec$ in the temporal network (\fref{fig:fig2}). When $b \sim 1/2$ (the numbers of events from $i$ to $j$ and from $j$ to $i$ are equal, i.e. the social tie is balanced), $\perec$ is large, meaning that the direction of the interaction between $i$ and $j$ changes repeatedly over time. Then, as $b$ moves away from $0.5$,  $\perec$ decreases, indicating that unidirectional interactions are more prevalent. Still, the fraction of reciprocations ranges from $0$ to the approximate upper bound $2(1-b)$ (for derivation see \srefsi{7}), meaning there is variability in $\perec$ among all datasets for a fixed value of $b$. Messaging, in particular, seems able to maximise reciprocity over balanced ties [i.e. $\perec \sim 2(1-b)$ for $b \sim 1/2$ in msg and sms]. Thus, $\perec$ complements balance as a measure of reciprocal relationships in communication networks, capturing its temporal nature more accurately.

Comparing datasets by $\perec$ for a given value of $b$ (\fref{fig:fig2}), we find that conversation channels show higher levels of reciprocity than Twitter. The datasets sms and msg show the largest reciprocity, followed by email and calls, with Twitter mentions and retweets at the lowest level of reciprocation (see \tref{tab_}). There are several potential explanations for this behavior. Short phone messages (sms) and direct messages within an online social network (msg) are usually directed at specific people and not used for broadcasting, meaning high reciprocity. Institutional communication (email) is often used both for sharing university-wide messages and talking among small groups of people, leading to heterogeneous values of reciprocity. Phone conversations (calls) are inherently bidirectional irrespective of who initiates the call, so people can be reciprocal within conversations even when data shows lower values of $\perec$. Twitter is consistently unidirectional mostly regardless of balance, in line with its use as a broadcast platform (see related results for in-/out-degrees in \srefsi{5}).

Human communication is typically bursty (made up of short trains of intense activity separated by long silences \cite{karsai2012correlated,unicomb2021dynamics}), making us wonder about the relationship between reciprocity and burstiness. We find, however, no significant correlation between $\perec$ and standard measures of burstiness \cite{barabasi2005origin,goh2008burstiness} (see \srefsi{6}). By separating communication channels into reciprocal and non-reciprocal temporal networks (see \fref{fig:fig1} top), we can also compute the time elapsed between successive reciprocations or non-reciprocations, which we refer to as the {\it time gap} $\Delta t$ (\fref{fig:fig3} left). The time gap is analogous to the well-known concept of inter-event time in temporal networks \cite{holme2012temporal,holme2015modern}, but between similar pairs of events ([non-]reciprocations) instead of single events.

The time gap distribution $p(\Delta t)$ shows that timescales of communication vary widely among channels -- sms has a fast dynamics with average time gap $\langle \Delta t \rangle \approx 0.5, 1.25$ days between successive reciprocations or non-reciprocations, respectively. Then we have calls, msg, mentions, retweets, and finally, emails as the slowest system with $\langle \Delta t \rangle \approx 28$ days between consecutive (non-)reciprocations. The broad distribution in the email channel seems to be consistent with its heterogeneous use for both sporadic institutional communication and more frequent personal exchanges. We also notice that reciprocation is faster than non-reciprocation in conversation channels (sms, msg, and email). The opposite is true for mentions, while calls and retweets show similar shapes of $p(\Delta t)$ between reciprocal and non-reciprocal exchange. Twitter as a broadcasting  platform shows more non-reciprocations and less time between them.

Following previous work on non-homogeneous patterns of communication activity over time \cite{barabasi2005origin,goh2008burstiness,karsai2012correlated,unicomb2021dynamics}, we extend the notion of burstiness to time gaps by defining $B = (\sigma - \mu) / (\sigma + \mu)$, where $\mu$ and $\sigma$ are, respectively, the mean and standard deviation of the time gaps between (non-)reciprocations. Time gap burstiness $B$ ranges  between -1 and +1, meaning time gaps are distributed either regularly or broadly in time. The difference between communication channels is even more evident when looking at the distribution $p(B)$ of time gap burstiness in both reciprocal and non-reciprocal components (\fref{fig:fig3} right). In conversation channels (sms, msg, email), reciprocal communication is significantly more bursty (i.e. less regular) than non-reciprocal exchange, while the broadcasting platform Twitter shows the opposite (non-reciprocity is more bursty). By explicitly separating communication into reciprocal and non-reciprocal components, sms comes out as the most non-homogeneous form of reciprocal communication among all channels considered. Overall,
the consideration of temporal reciprocity, time gaps, and burstiness allow us to identify a spectrum of roles of communication (from one-on-one communication to uni-directional broadcast) not apparent from aggregated data alone.

\begin{table*}[t]
\begin{tabular}{!{\vrule width 1pt}l!{\vrule width 1pt}c|c|c|c!{\vrule width 1pt} c|c|c|c!{\vrule width 1pt}c|c|c|c!{\vrule width 1pt}}
\hline
\toprule
\rowcolor[HTML]{FFFFFF}
    &   \multicolumn{4}{c!{\vrule width 1pt}}{$\perec$}&   \multicolumn{4}{c!{\vrule width 1pt}}{$\plrec$}\\
\rowcolor[HTML]{FFFFFF}
\hline
 \textbf{Dataset \textbackslash Method}                  & NTS & NDS & NTSR & NDSR   & NTS & NDS & NTSR & NDSR  \\
\hline
 \rowcolor[HTML]{FFFFFF} 
sms &  + $\blacksquare$ & +  $\blacksquare$& +  $\blacksquare$& +  $\blacksquare$     & \hspace{2.4 mm} \scalebox{1.45}{$\circ$} & \hspace{0.1 mm} - \scalebox{0.95}{$\triangle$}  & + \scalebox{1.15}{$\blacktriangle$} & + \scalebox{0.95}{$\triangle$}   \\
 \rowcolor[HTML]{FFFFFF}%{FFFFFF}
msg    &   + $\blacksquare$  & +  $\blacksquare$ & + $\blacksquare$  & +  $\blacksquare$    &  \hspace{2.4 mm} \scalebox{1.45}{$\circ$} & + \scalebox{1.15}{$\blacktriangle$} & + \scalebox{1.15}{$\blacktriangle$}  & \hspace{0.1 mm} - $\blacksquare$    \\
 \rowcolor[HTML]{FFFFFF}
email  & + $\blacksquare$  & + $\blacksquare$ & + $\blacksquare$  & + \scalebox{0.95}{$\triangle$}  &	+  $\blacksquare$ & + \scalebox{1.15}{$\blacktriangle$}  & + \scalebox{1.15}{$\blacktriangle$}& \hspace{0.1 mm} - $\blacksquare$  \\
\rowcolor[HTML]{FFFFFF}
retweets  & \hspace{0.1 mm} - $\blacksquare$  & \hspace{0.1 mm} - $\blacksquare$  & \hspace{0.1 mm} - \scalebox{1.15}{$\blacktriangle$}  & \hspace{0.1 mm} - \scalebox{1.15}{$\blacktriangle$} & \hspace{2.4 mm} \scalebox{1.5}{$\circ$}  & +  $\blacksquare$  & \hspace{0.1 mm} - \scalebox{1.15}{$\blacktriangle$} & \hspace{0.1 mm} - \scalebox{1.15}{$\blacktriangle$}     \\
\rowcolor[HTML]{FFFFFF}
mentions  & \hspace{0.1 mm} - $\blacksquare$  &  \hspace{0.1 mm} - $\blacksquare$ & \hspace{0.1 mm} - $\blacksquare$ & \hspace{0.1 mm} - $\blacksquare$ & + $\blacksquare$ &  + \scalebox{0.95}{$\triangle$} & +  $\blacksquare$ &  \hspace{0.1 mm} - $\blacksquare$   \\ \hline
\end{tabular}
\caption{\textbf{Null models identify memory as mechanism for reciprocity.} Sign and order of magnitude of $z$-scores when comparing the reciprocity measures $\perec$ and $\plrec$ between the studied datasets and four null models shuffling interaction events. Symbols are $\circ$ ($z = 0$), $\triangle$ ($|z| < 2 $), $\blacktriangle$ ($2 <|z| < 10 $), and $\blacksquare$ ($10 <|z| < 100 $), with filled symbols indicating statistical significance (i.e. large magnitude). A negative, close to zero, or positive $z$-score implies that the null model over-estimates, captures, or under-estimates the empirical measure, respectively. Null models are denoted by NTS (Network Shuffling Timestamps), NDS (Node Shuffling Timestamps), NTSR (Network Rewiring and Shuffling Timestamps), and NDSR (Node Rewiring and Shuffling Timestamps) (see \srefsi{4}). The calls dataset is not included due to its small size after filtering (see \srefsi{3}). Overall we see more positive than negative $z$-scores, implying that reciprocity is not reproduced by random mechanisms, and suggesting memory as a relevant mechanism for reciprocal interaction in social communication. There is also a notable difference in $z$-score sign between conversation (sms, msg, email) and broadcasting (retweets, mentions) channels, pointing to the distinct roles of bidirectional vs. unidirectional exchange.
}
 \label{Shuffling_table}
\end{table*}

\subsection*{Null models identify memory as mechanism for reciprocity}

Having established the presence of reciprocity and its temporal features in several communication channels, we turn to the question of how much of the reciprocation seen in data is explained simply by random processes, and how much is otherwise potentially due to specific mechanisms of social interaction, particularly memory \cite{karsai2014time}. In line with previous work on random models of reciprocity in static networks \cite{garlaschelli2006multispecies,zamora2008reciprocity,zlatic2009influence,squartini2013reciprocity}, we focus on four null models that randomize (i.e. shuffle) the time of occurrence of events and/or the network topology. As a task of hypothesis testing via reference models of temporal networks \cite{gauvin2018randomized}, our null models correspond to the class of microcanonical randomized reference models, since we impose constraints on some network features (e.g., degree, number of events, etc.), while randomly shuffling others (e.g., time ordering of events, links, etc.).

The null models considered include two types of shuffling: a) {\it timestamp shuffling}, or b) {\it rewiring and timestamp shuffling}. Timestamp shuffling keeps the network topology fixed while randomly exchanging the times of event occurrence, thus randomizing the temporal aspects of communication only, not the underlying pattern of interactions. The rewiring and timestamp shuffling method randomizes both the network topology and timestamps of event occurrence, affecting temporal and structural patterns of information exchange. We implement the two shuffling methods at two levels of resolution: a) {\it node level} or b) {\it network level}. Shuffling at the node level is applied to the ego networks of each node independently, while shuffling at the network level is applied to all nodes at once. The combination of a shuffling method and a level leads to four null models, which we denote: (i) NTS (Network Shuffling Timestamps), (ii) NDS (Node Shuffling Timestamps), (iii) NTSR (Network Rewiring and Shuffling Timestamps), and (iv) NDSR (Node Rewiring and Shuffling Timestamps) (for a detailed description of each null model see \srefsi{4}).

In line with the observation that humans remember past contacts and often repeat them over time \cite{karsai2014time}, the analysis of our null models suggests memory is an underlying mechanism for reciprocal interactions (\tref{Shuffling_table}). Particularly, the null model NTSR preserves the in- and out-degree of each node in the network while randomizing the identity of the nearest neighbors, as well as the times of occurrence of their events. By disregarding previous social contacts or their event times, this null model erases the memory of agents, both in the structural and temporal sense. We quantify this effect by calculating $z$-scores, i.e. the difference between the value of a measure in data and its average over an ensemble of realizations of the null model, relative to the standard deviation of the measure in the ensemble.  Both proposed measures of reciprocity [$\perec$ and $\plrec$] show large and positive $z$-scores, especially for NTSR, indicating that empirical communication channels have more reciprocation than the randomized reference model. This lack of reciprocation upon removing memory mechanisms, while preserving individual and network properties, suggests memory as an relevant driver for reciprocity in social communication networks. In the other three null models (NTS, NDS, and NDSR), we see a similar trend in $z$-scores for $\perec$, while  $z$-scores for $\plrec$ are somewhat similar for each separate null model, across all datasets. These results indicate that $\perec$ is a useful measure of actual reciprocity in the network, in the sense that it reacts in similar ways to random events and noise from system to system.

A comparison of the values of $\perec$ between empirical data and the null models also highlights the distinct roles of more traditional communication channels (sms, msg, and email, mostly used for one-on-one conversations) as opposed to the broadcasting platform Twitter (retweets, mentions) (\tref{Shuffling_table}). Conversation channels all have positive and large $z$-scores, meaning that empirical values of $\perec$ are higher than their randomized counterparts in all shuffling methods, while the opposite happens in Twitter. We interpret this behaviour as an increased tendency for reciprocal and bursty interactions in conversation channels. Communication in Twitter seems less reciprocal and bursty, possibly due to the intended use of the platform as a public setting dominated by unidirectional messaging aimed towards wider audiences.

%\todo{Luka: Briefly discuss somewhere the results of other null models, apart from NTSR}

\subsection*{Modeling reciprocity in temporal networks}

Efforts at theoretically understanding the emergence of reciprocal interactions in temporal communication data include Bayesian inference via network models of Hawkes processes \cite{blundell2012modelling,miscouridou2018modelling} and stochastic blockmodeling of relational event data \cite{dubois2013stochastic} in both directed \cite{safdari2021generative} and temporal \cite{safdari2022reciprocity} networks. When posed as a machine learning task, the identification of reciprocal interactions has also been applied to the prediction of online extremism in Twitter \cite{ferrara2016predicting}. Here, we attempt to model the temporal patterns of reciprocity seen in empirical data via a flexible framework of activity-driven (AD) temporal networks \cite{perra2012activity}, used previously to explore several features of human communication dynamics, from cognitive constraints \cite{gonccalves2011modeling} to social contagion \cite{liu2014controlling}.

The AD model introduces a (typically broad) activity potential to describe the dynamics of structural heterogeneity in temporal networks \cite{perra2012activity}: active nodes are chosen more frequently to interact with other randomly selected nodes, with no memory of past interactions. Empirical communication data shows, however, a tendency of individuals to communicate preferentially over established social connections. Indeed, a previous analysis of mobile call networks \cite{karsai2014time} shows that, as time goes by and social circles evolve, individuals are more likely to re-contact someone they already know, and less likely to interact with new people. Ref. \cite{karsai2014time} extends the AD model to include a notion of {\it memory} (the ADM model), which promotes connections with past neighbours. Independently, the AD model has also been extended with a concept of {\it attractiveness} (the ADA model), by which an individual aggregates more incoming connections from active nodes than from others \cite{pozzana2017epidemic}.

Here we combine both features (attractiveness and memory) into a single model, ADAM, and use it to reproduce the observed levels of reciprocity in our six datasets. We define the activity $a_i$ and attractiveness $b_i$ of node $i$ as
\begin{equation}
    a_i= \frac{\sum_t{k}_{out}(i,t)}{\sum_{\ell,t} {k}_{out}(\ell,t)}, \hspace{3pt}
    b_i= \frac{\sum_t{k}_{in}(i,t) }{\sum_{\ell,t}{k}_{in}(\ell,t)},
\end{equation}
where $k_{in}(i,t)$ and $k_{out}(i,t)$ are the empirical in- and out-degrees of node $i$ at time $t$. In other words, the activation probability is proportional to out-degree and the attractiveness to in-degree. Then, the ADAM model follows the next rules recursively:
\begin{itemize}
    \item At each discrete time step $t$ the synthetic network starts with $N$ disconnected nodes.
    \item With probability $a_i \Delta t$ each node $i$ becomes an active source node and generates $m$ out-stubs (or half-links). For each out-stub,
\begin{itemize}
    \item[-] \emph{(memory step)} with probability $c/(c+k)$, where $c$ is a memory parameter, select target node $j$ from the past contacts of node $i$, according to its attractiveness $b_j$. The memory parameter $c$ is fitted from each dataset as in \cite{karsai2014time}. 
    \item[-] Otherwise, the target $j$ is chosen randomly from the whole population with probability equal to its attractiveness $b_j$. 
\end{itemize}

    \item At the next time step $t+\Delta t$, all edges in the synthetic network are deleted. Thus, all interactions have a constant duration $\Delta t$.
\end{itemize}

\begin{figure}[t]
    \centering
\includegraphics[width =0.95\columnwidth]{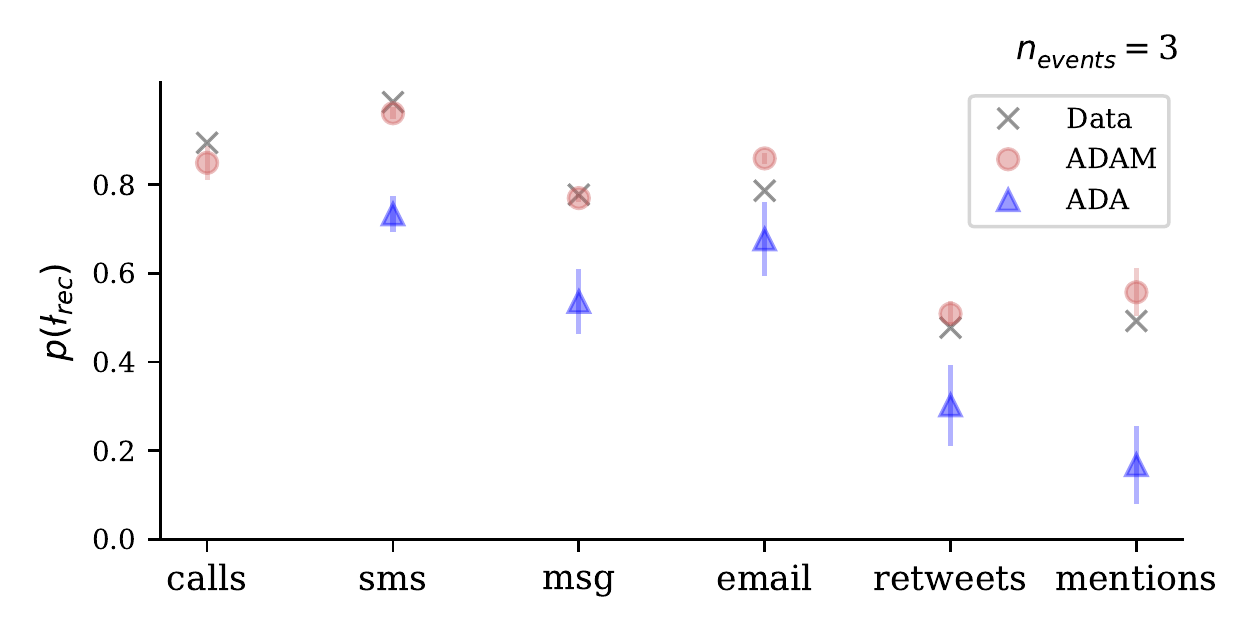}
\includegraphics[width =0.95\columnwidth]{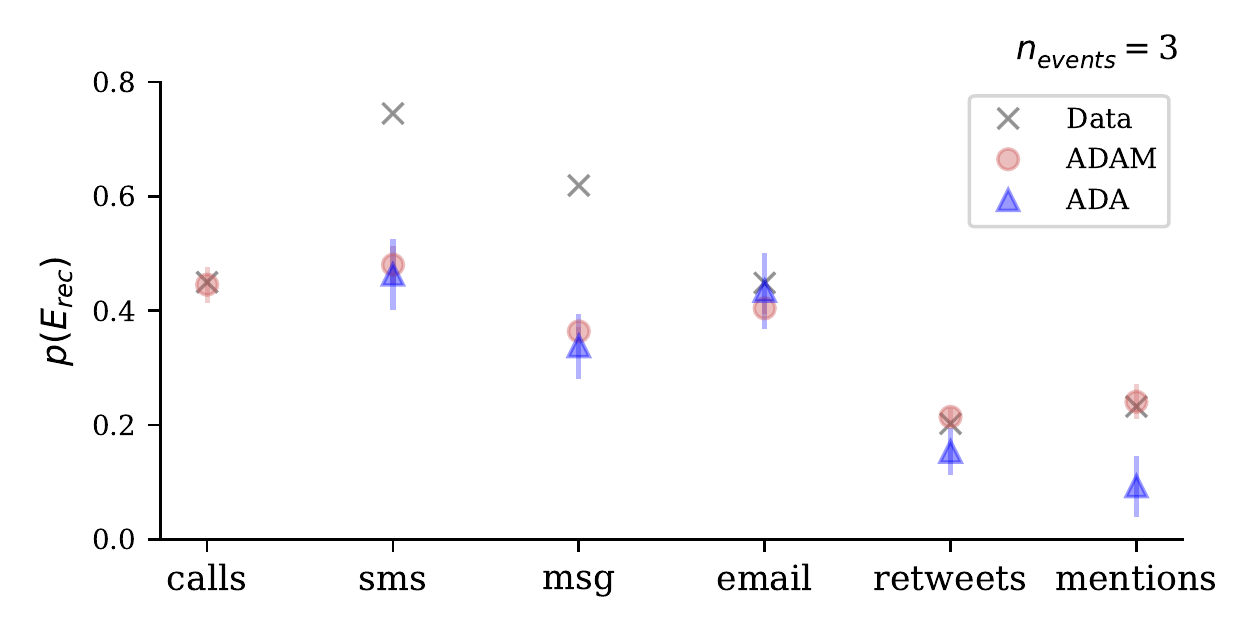}
\caption{\textbf{Memory drives reciprocity in activity-driven network model.} Fraction of links having at least one reciprocation, $\plrec$ (top), and fraction of reciprocations $\perec$ (bottom), for several empirical communication channels (Data), as well as in synthetic temporal networks fitted by the ADAM and ADA models. The ADAM model, implementing both memory and attractiveness of nodes, reproduces empirical levels of reciprocity in all datasets, outperforming the memoryless ADA model.  We filter out links with less than $n_{events}=3$ events. For calls, ADA does not produce any links with more than 3 events on them, so all edges are filtered out and the ADA symbol disappears. Qualitatively similar results are obtained for a different choice of filter, $n_{events}=5$ (see \srefsi{8}).}
\label{fig:fig4}
\end{figure}

We numerically simulate the ADAM model, produce synthetic temporal communication networks and measure levels of reciprocity via $\plrec$ and $\perec$ (\fref{fig:fig4}). Comparison against an ADA model (i.e. lacking memory) serves as a baseline for testing the performance of our model. The ADAM model captures very well $\plrec$, consistently outperforming ADA model across all channels considered. Values of $\perec$ are well reproduced by ADAM for retweets, mentions and calls, while ADA fails for all but email. Overall, a preference to preferentially interact with active individuals and previous social contacts, both within an activity-driven framework, seems enough to reproduce the temporal patterns of reciprocity observed in several communication channels.

Note that ADAM is not able to reproduce $\perec$ for sms and msg, perhaps due to a more complex role of memory in these communication channels. The ADAM model does indeed account for memory of past contacts; however, it ignores the possibility that alters are treated differently by an ego. Namely, strong weight heterogeneity over the links of aggregated ego networks might cause discrepancies between data and ADAM. In any case, ADAM outperforms ADA even in the case of sms and msg, showcasing the relevance of some type of memory effect. Future research in the drivers of reciprocity in social communication networks might consider more involved implementations of memory, such as one where the number of past contacts with a given individual determines the frequency of future interactions.

\section*{Discussion}

In this paper, we have proposed measures of reciprocity that explicitly account for the temporality of social interactions in human communication, and used them to quantity the levels of reciprocation in multiple channels including calls, messaging and social media. We have shown that existing reciprocity measures on aggregated directed and weighted networks \cite{garlaschelli2004patterns,squartini2013reciprocity}, particularly the notion of balance \cite{karsai2012correlated}, are actually an upper bound on temporal reciprocity measures like $\perec$. Given a level of balance between pairs of nodes, temporal reciprocity can vary widely, highlighting differences across communication channels. Indeed, for conversation channels like sms, msg and email, the time gaps between successive reciprocations tend to be shorter than between successive non-reciprocations. This suggests that one-on-one channels \cite{jensen2011internet} support quicker reciprocal communication than the broadcasting platform Twitter. We have seen a similar effect for time gap burstiness; conversation channels have more bursty reciprocal activity, while Twitter displays more bursty non-reciprocal dynamics. Implementing several null models based on event shuffling \cite{gauvin2018randomized}, we have identified the memory of past contacts as a driver of reciprocity. Upon adding a memory mechanism to a framework of activity-driven temporal networks \cite{perra2012activity,karsai2014time,pozzana2017epidemic}, we were also able to theoretically emulate the observed levels of reciprocity in several communication channels. 

Even if more granular than previous measures on aggregate data, the quantities $\perec$ and $\plrec$ are themselves upper bounds on reciprocal activity over a social tie. We define reciprocity as pairs of events in opposite directions across the tie, regardless of the time elapsed between them. But if this time is too large, events are potentially not related to each other (e.g., correspond to different conversations, topics, or even people), meaning actual reciprocity is equal or lower than $\perec$ and $\plrec$. This effect might not be large given our observation that reciprocal communication is bursty, i.e. trains of reciprocation with small time gaps between them are common, notably in conversation channels (\fref{fig:fig3}). Still, it remains an open question whether our measures could be extended beyond event directionality to reflect reciprocal human behaviour more closely, by, for example, integrating temporal correlations, or communication content via text analysis \cite{esau2022creates}.

Our exploration of patterns of reciprocity in human communication deals with the large-scale structure of temporal networks. We identify reciprocal interactions at the link level and then accumulate them over whole channels. This reveals a spectrum of modes of communication, from reciprocal, one-on-one conversation channels, to non-reciprocal platforms used mainly for broadcasting. The way reciprocity is distributed across the ego network of an individual is, however, still unexplored. Social signatures, a ranking of alters by decreasing number of contacts with the ego, seem to persist in time and across communication channels \cite{saramaki2014persistence,heydari2018multichannel} and correlate with individual traits \cite{aledavood2016channel}. Alter turnover also grows as we go down the ranking, in agreement with generic behavior of rankings in social systems \cite{iniguez2022dynamics}. By extending our measures to the dynamics of social signatures, we might find higher levels of reciprocal activity among top alters, further cementing the relationship between reciprocity and notions of stability and cohesion in social networks.

%%%%%%%%%%%%%%%%%%%%%%%%%%%%%%%%%%%%%%%%%%%%%%
%%                                          %%
%% Backmatter begins here                   %%
%%                                          %%
%%%%%%%%%%%%%%%%%%%%%%%%%%%%%%%%%%%%%%%%%%%%%%

\begin{backmatter}

\section*{Availability of data and materials}

Code to reproduce the results of the paper is publicly available at \href{https://github.com/dynamicalsystemsceu/codes}{github.com/dynamicalsystemsceu/codes}.
For data availability see \srefsi{2}. Non-public data is available from the authors upon reasonable request.

\section*{Competing interests}

The authors declare that they have no competing interests.

\section*{Funding}

G.I.  acknowledges support from AFOSR (Grant No. FA8655-20-1-7020),  project EU H2020 Humane AI-net (Grant No. 952026),  and CHIST-ERA project SAI (Grant No. FWF I 5205-N). 

\section*{Authors' contributions}

S.C., E.A., A.M., and L.B. contributed equally to this work, and are listed in order agreed by the authors. E.A. and A.M. analyzed the empirical data, and A.M. crawled the Twitter data. L.B. implemented the null models, and S.C. explored the reciprocity model. All authors conceived the research, discussed the results, and wrote the manuscript.

\section*{Acknowledgements}

This project was developed as part of the `Dynamical systems in networks' course taught by G.I. at the Department of Network and Data Science in Central European University.  We acknowledge the participation of Pavel Kiparisov in the initial stages of the project.
  
%%  After submission of the .TEX file,                     %%
%%  you will be prompted to submit your .BBL file.         %%

% if your bibliography is in bibtex format, use those commands:
\bibliographystyle{bmc-mathphys} % Style BST file (bmc-mathphys, vancouver, spbasic).
\bibliography{refs}      % Bibliography file (usually '*.bib' )
% for author-year bibliography (bmc-mathphys or spbasic)
% a) write to bib file (bmc-mathphys only)
% @settings{label, options="nameyear"}

% or include bibliography directly:
% \begin{thebibliography}
% \bibitem{b1}
% \end{thebibliography}

\end{backmatter}
\end{document}

\begin{center}
{\LARGE Supplementary Information for}\\[0.7cm]
{\Large \textbf{Temporal patterns of reciprocity in communication networks}}\\[0.5cm]
{\large S. Chowdhary, E. Andres, A. Manna, L. Blagojevic, L. Di Gaetano, G. I\~niguez$^*$}\\[0.7cm]
{\small $^*$Corresponding author email: iniguezg@ceu.edu}\\[2cm]
\end{center}

\section{Temporal network concepts and reciprocity measures}

\textbf{Events}\\
An \textit{event} $e_{ijt}$ is the interaction (communication) between source node $i$ and target node $j$ at time $t$.

\noindent
\textbf{Link}\\
A link $l_{ij}$ exists between nodes $i$ and $j$ if at least one event happens between them in the observation period, i.e. in the underlying static aggregate network (an undirected, simple network where link weights are discarded).

\noindent
\textbf{Node event sequence}\\
A sequence of events of node $i$ is, e.g., $S_i =  \{e_{ijt_1},e_{kit_2},e_{imt_3}...e_{jit_T}\}$, that is, a series of $T$ events where $i$ is always involved, either as target or source node. Events are then out-links or in-links happening at some time $t$, i.e. the communication interactions that an ego has with its alters. 

\noindent
\textbf{Link event sequence}\\
A sequence of events in link $l_{ij}$ between nodes $i$ and $j$ is, e.g., $S_{ij} =\{e_{ijt_1},e_{jit_2},e_{ijt_3}...e_{jit_T}\}$, that is, a series of $T$ events involving $i$ and $j$ in any direction. Events are then out-links or in-links happening at some time $t$, i.e. the communication interactions between the pair of individuals. 

\noindent
\textbf{Reciprocal events}\\
A pair of events is reciprocal if events are consecutive and the direction of the second event is opposite to the first, i.e. $(e_{ijt_1}, e_{jit_2})$ where $t_2 > t_1$.

\noindent
\textbf{Reciprocal link}\\
A reciprocal link contains at least one reciprocation (reciprocal event pair) in its sequence of events. We compute: (i) the number of reciprocations $E_{rec,ij}$ over link $l_{ij}$, relative to the number of consecutive event pairs on that link, $E_{ij}-1$. By averaging over links, we obtain the reciprocation probability $\perec = \langle E_{rec,ij} / (E_{ij}-1)\rangle_{ij}$. We also compute: (ii) the number of links with at least one reciprocation ($l_{rec}$) relative to the total number of links ($L$), $\plrec = l_{rec} / L$.

\noindent
\textbf{Inter-event time}\\
Inter-event time is the time span between consecutive events in a sequence. Inter-event times can be computed for both node and link event sequences.

\noindent
\textbf{Time gap}\\
A time gap is the time elapsed between two successive reciprocations or non-reciprocations. It is analogous to an inter-event time, but between event pairs instead of single events.

\noindent
\textbf{Time gap burstiness}\\
Time gap burstiness is defined as $B = (\sigma - \mu) / (\sigma + \mu)$, where $\mu$ and $\sigma$ are, respectively, the mean and standard deviation of the time gaps between (non-)reciprocations. Time gap burstiness $B$ ranges  between -1 and +1, meaning time gaps are distributed either regularly or broadly in time. It can be computed for both node and link event sequences.

\section{Data description}
\label{sec:data}

We analyze several datasets of social contact between individuals from a wide range of studies in the temporal networks literature (see Table 1 in main text). Each dataset includes a time-ordered set of communication events between anonymized individuals $i$ and $j$ (according to hashed timestamps). From a dataset we construct a temporal network where a directed link $l_{ij}$ appears instantaneously if individual $i$ initiates an event towards individual $j$ at some point in time.

\subsection{List of datasets}
\label{sec:dataList}

\paragraph{Copenhagen Networks Study (calls \& sms).} Dataset of multi-channel, phone-enabled social interactions from the Copenhagen Networks Study (CNS) \cite{stopczynski2014measuring,sapiezynski2019interaction}. The original study includes activity of roughly 1,000 individuals during 2012-2013 \cite{stopczynski2014measuring}. Data used here is a selected portion of the full dataset as described in \cite{sapiezynski2019interaction}. The dataset includes events in two channels (disregarding Bluetooh data): call and short message logs between individuals, with data on timestamps of the call/message, anonymized user IDs, and call duration. We also delete missed calls, making the dataset smaller from the one in \cite{sapiezynski2019interaction}. Data is publicly available via {\it figshare} in \cite{data_availability}.

\paragraph{Twitter (retweets \& mentions).} The Twitter dataset has been collected through the Twitter API from 2018-02-18 to 2018-12-18. The request was limited to tweets that contained at least one selected keyword (*vaccin*\footnote{The string \textit{`{*vaccin*}'} allows us to capture every possible declination and compounding of the Italian word \textit{vaccino} (vaccine) and the verb \textit{vaccinare} (vaccinate).}, vax, libertàdiscelta, libertadiscelta, ddl770, trivalente, \#mmr), with no limitation to the location nor to the language of the tweet. From the list of tweets collected, two directed temporal networks have been created according to the type of action (retweets and mentions), and we disregard replies. In all Twitter networks, nodes represent users, while a link goes from $i$ to $j$ if user $i$ mentioned  $j$ or retweeted a post of user $j$.

\paragraph{College messages (msg).}
This dataset is comprised of private messages sent on an online social network at the University of California, Irvine \cite{panzarasa2009patterns} from April to October 2004. Users join an online community, intended to help students communicate with their friends, and to meet new people. First, members have to create an account by filling in some information; then they can search the network for others and then initiate conversation based on profile information. There is a total of 1,899 users, who have exchanged 59,835 online messages.

\paragraph{EU research institution (email).}
The network was generated using email data from a large European research institution \cite{paranjape2017motifs} from October 2003 to May 2005. It comprises 3038531 emails, sent from 287755 different email addresses. All information has been anonymized. Emails only represent communication between institution members (the core), and the dataset does not contain incoming messages from or outgoing messages to the rest of the world.

\section{Event filtering}

To ensure statistically significant results, we filtered the original networks in such a way that each remaining link has a minimum number of events. This decision is mainly motivated by previous works \cite{granovetter1973strength,marsden1984measuring,wang2013dyadic}, where the authors claim that the quantity of interactions between people is a good proxy for social tie strength. We then focus on significant ties, meaning that the two involved individuals have exchanged some minimum number of messages (events) between themselves. 

\begin{figure}[t]
    \centering
    \begin{minipage}{0.45\textwidth}
        \centering
        \includegraphics[width=0.9\textwidth]{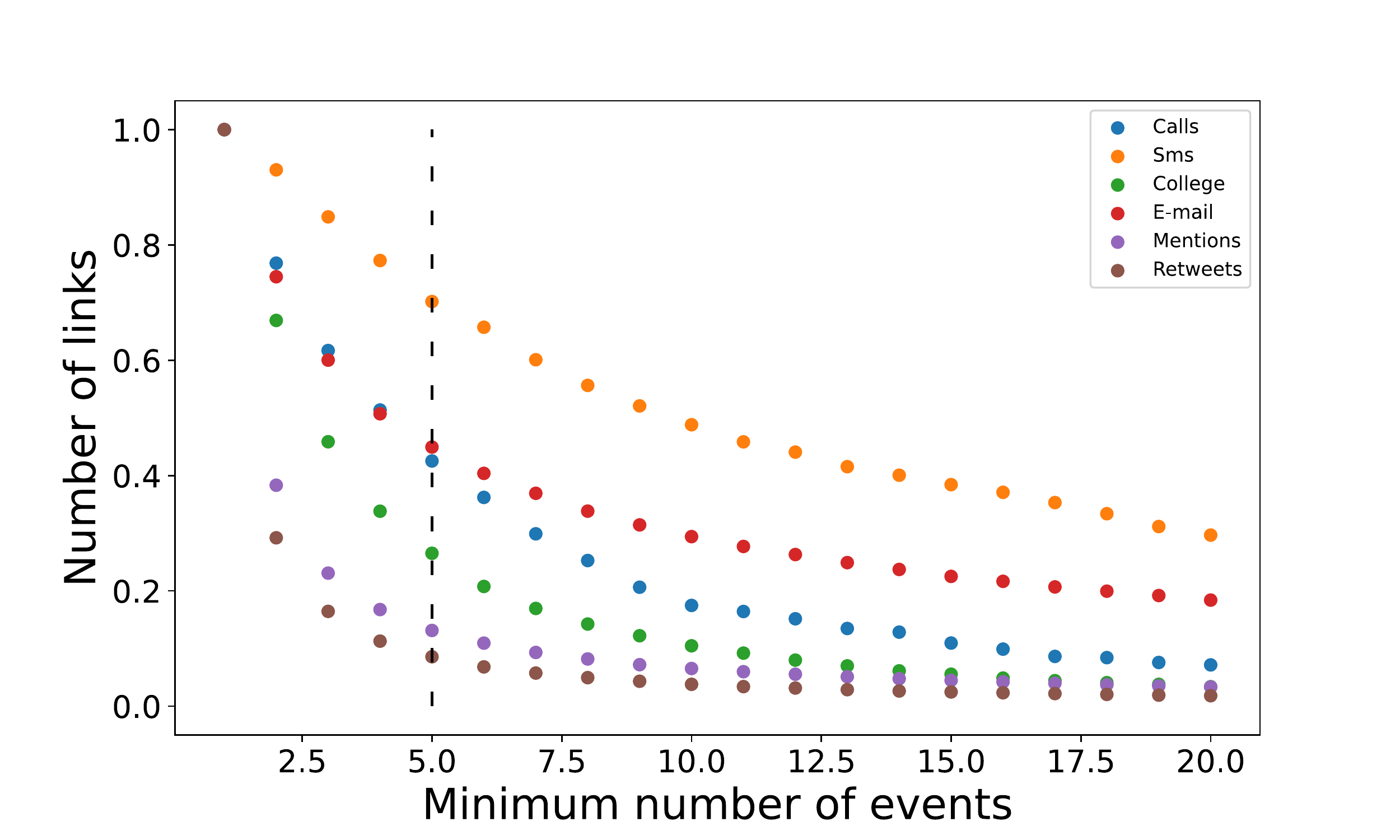} % first figure itself
        %\caption{Raw Data}
    \end{minipage}
    \begin{minipage}{0.45\textwidth}
        \centering
        \includegraphics[width=0.9\textwidth]{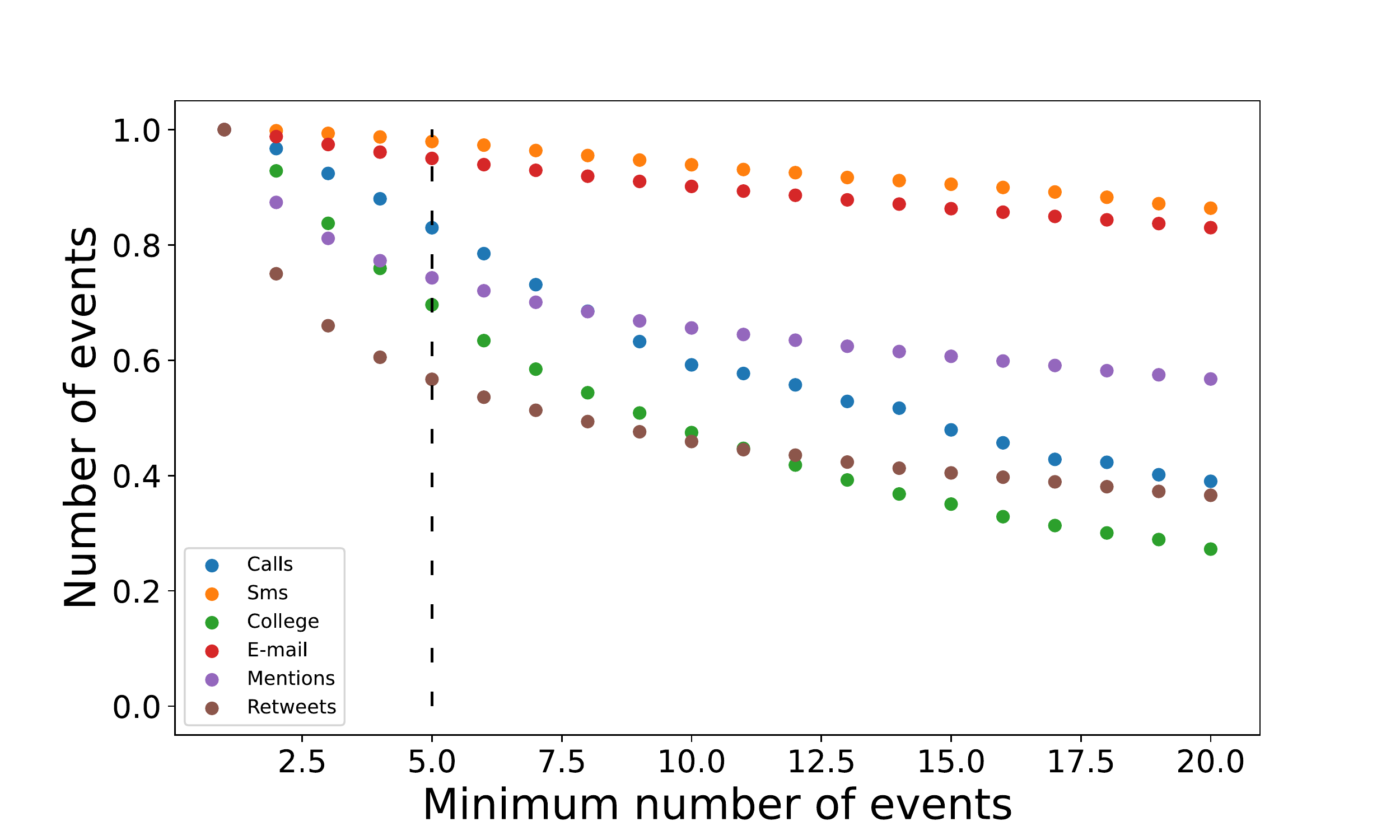} % second figure itself
        %\caption{Reconstructed data that uses points to include local thickness}
    \end{minipage}
    \begin{minipage}{0.45\textwidth}
        \centering
        \includegraphics[width=0.9\textwidth]{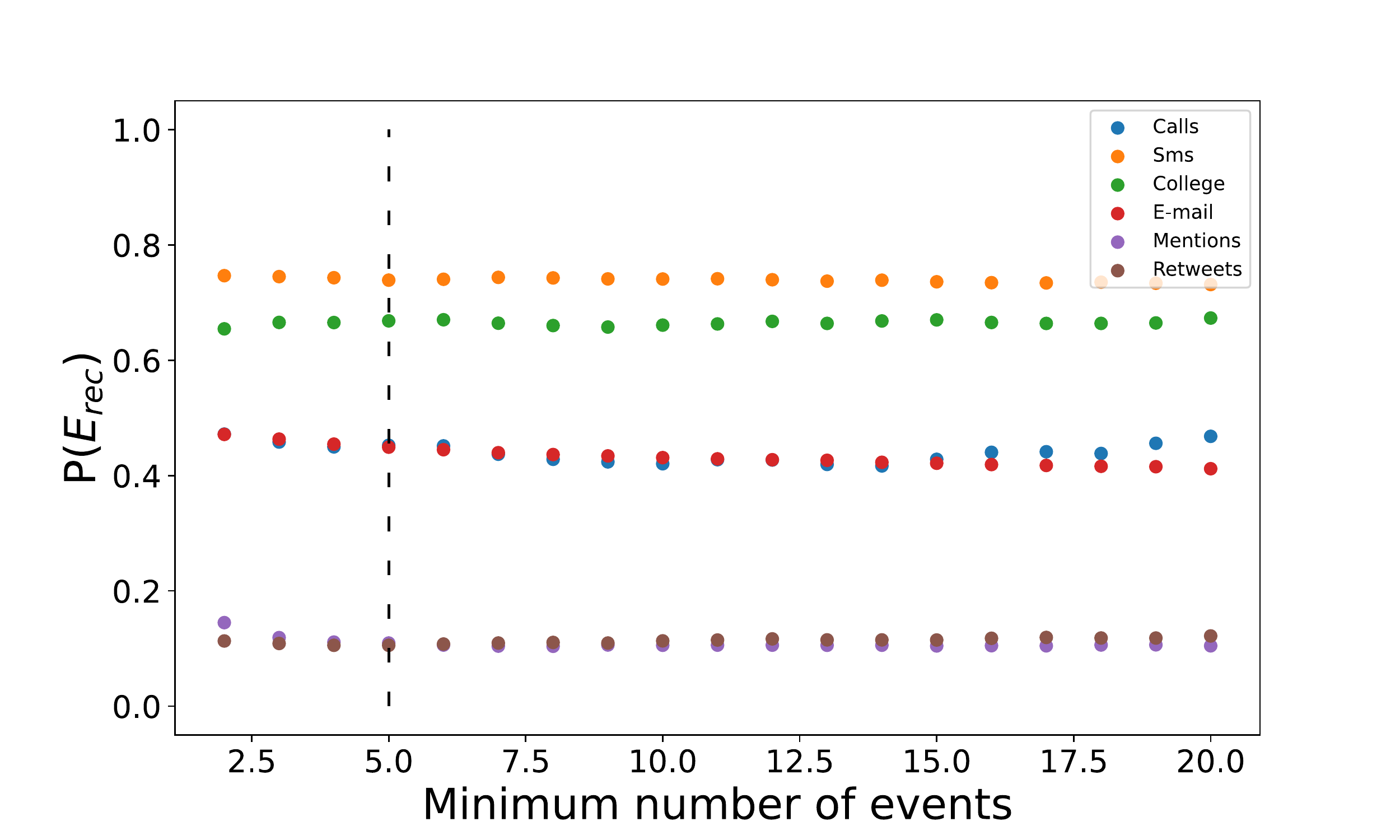} % second figure itself
        %\caption{Reconstructed data that uses points to include local thickness}
    \end{minipage}
    \begin{minipage}{0.45\textwidth}
        \centering
        \includegraphics[width=0.9\textwidth]{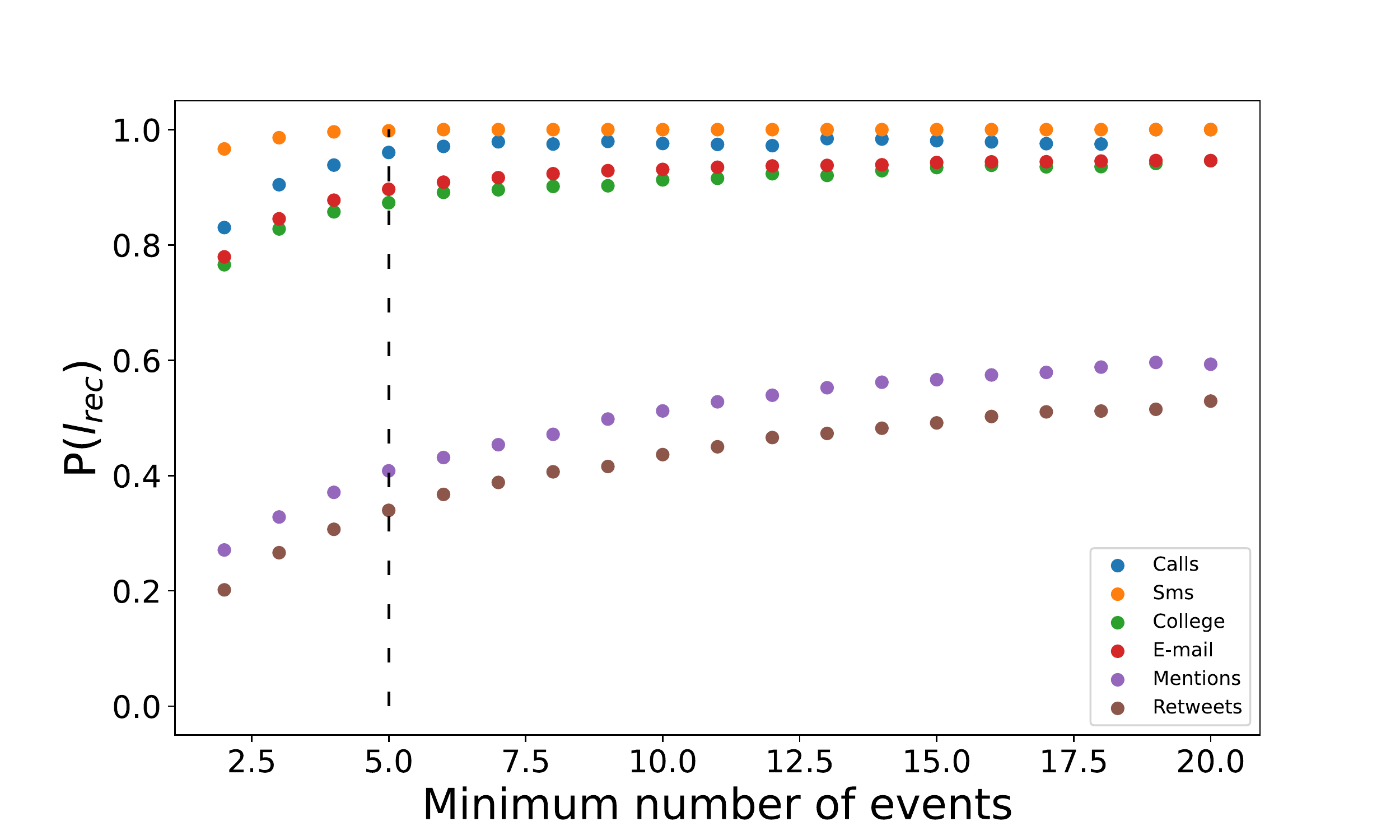}% second figure itself
        %\caption{Reconstructed data that uses points to include local thickness}
    \end{minipage}
    \caption{Network and reciprocity measures as a function of filtering parameter. The dashed line represents the chosen value for the filter, where there is relative stability across all datasets.}
    \label{fig:fig_filtering}
\end{figure}

We have to pick a value equal to or higher than 3, since we need at least 2 inter-event times to compute their standard deviation, which we use in our exploratory results. Based on our sensitivity analysis, where we varied the minimum number of events per link (\fref{fig:fig_filtering}), we fix this filtering parameter to $5$, marked by the vertical dashed line. This is the minimum value for which most measures stabilize, meaning that their rate of change is relatively low, compared to the smaller values of other filtering parameter values (3 or 4 minimum events per link).

\section{Null models}

We utilize null models to assess how much our measures differ when computed in the original empirical networks, compared to the their randomized versions. By combining two shuffling types and two resolution levels, we employ four different null models (randomized reference models):
\begin{enumerate}
    \item[-] \textbf{NTS (Network Shuffling Timestamps)}. All events occur at their original links, between the same nodes, while each time of occurrence for every event is sampled without replacement from the set of all times of occurrence for the network. This means that all events occur between the same nodes, but each of them at a different, randomly sampled time.
    \item[-] \textbf{NDS (Node Shuffling Timestamps)}. All events occur at their original links, between the same nodes, but each event randomly obtains a time of occurrence sampled with a replacement from the set of events that belong to its neighboring nodes. This means that all events occur between the same nodes, but at different times, randomly determined from their initial neighborhood.
    \item[-] \textbf{NTSR (Network Rewiring and Shuffling Timestamps)} Network links are randomly reassigned, with in- and out-degrees mostly conserved for each node (unless self-loops occur, which happens in the configuration model). Additionally, the time of occurrence of each event is randomly sampled without replacement from the set of all events in the network. This means that the same number of events occurs, but between randomly sampled nodes within a network and at randomly sampled times.
    \item[-] \textbf{NDSR (Node Rewiring and Shuffling Timestamps)}  All links are randomly reassigned, with in- and out-degrees conserved for each node, and each node randomly obtains a link sampled with replacement from the set of its initial neighbors. Additionally, the time of occurrence for each event is randomly sampled without replacement from the set of all neighbor's events. This means that the same number of events occurs, but between randomly sampled nodes from their original neighborhoods and at times randomly sampled from their neighbors.
\end{enumerate}

All shuffling methods are iterated 50 times for all datasets, except for the Twitter retweets dataset where 39 iterations are performed (due to larger size of the dataset). By computing mean values and standard deviations for the obtained distribution of measures, we obtain $z$-scores for each dataset (see Table 2 in main text).

\begin{figure}[t]
\centering
\includegraphics[scale=0.5]{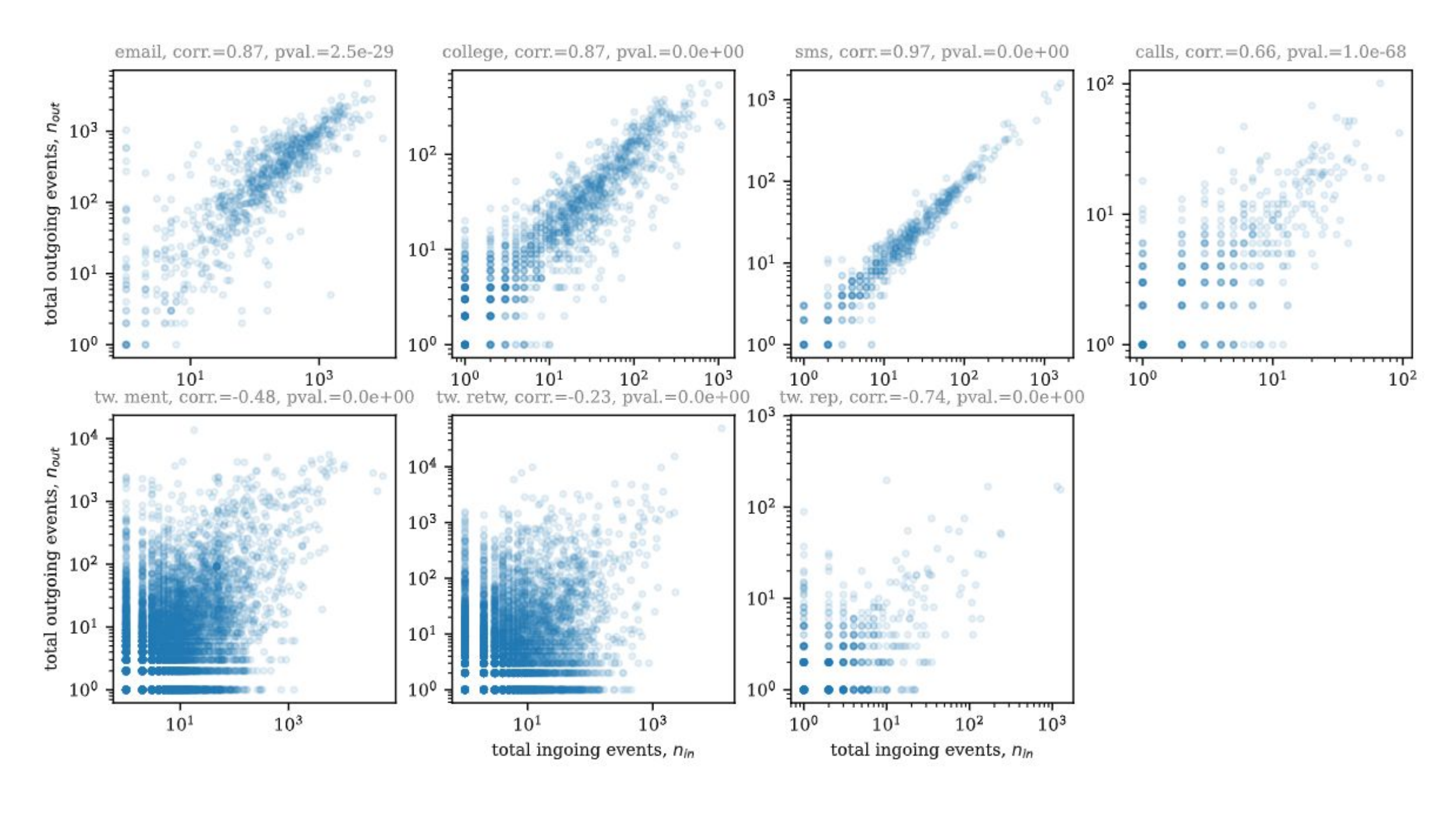}
\caption{Scatter plot of in- and out-degrees of all networks before filtering. We show the associated Pearson correlation coefficients (corr) and $p$-values (pval).}
\label{inout}
\end{figure}

\section{Correlation between in and out degrees}

\fref{inout} shows a scatter plot of in-degrees and out-degrees in each of the studied networks. For all of them, we compute the Pearson correlation coefficient, which shows positive and significant correlations between these two quantities, thus revealing the presence of reciprocity relationships. Apparently, this correlation is higher for conversation channels than for Twitter, further highlighting the use of Twitter as a broadcasting platform, not a place for one-on-one communication.

\section{Correlation between standard burstiness and reciprocity}

\begin{figure}[t]
\centering
\includegraphics[scale=0.5]{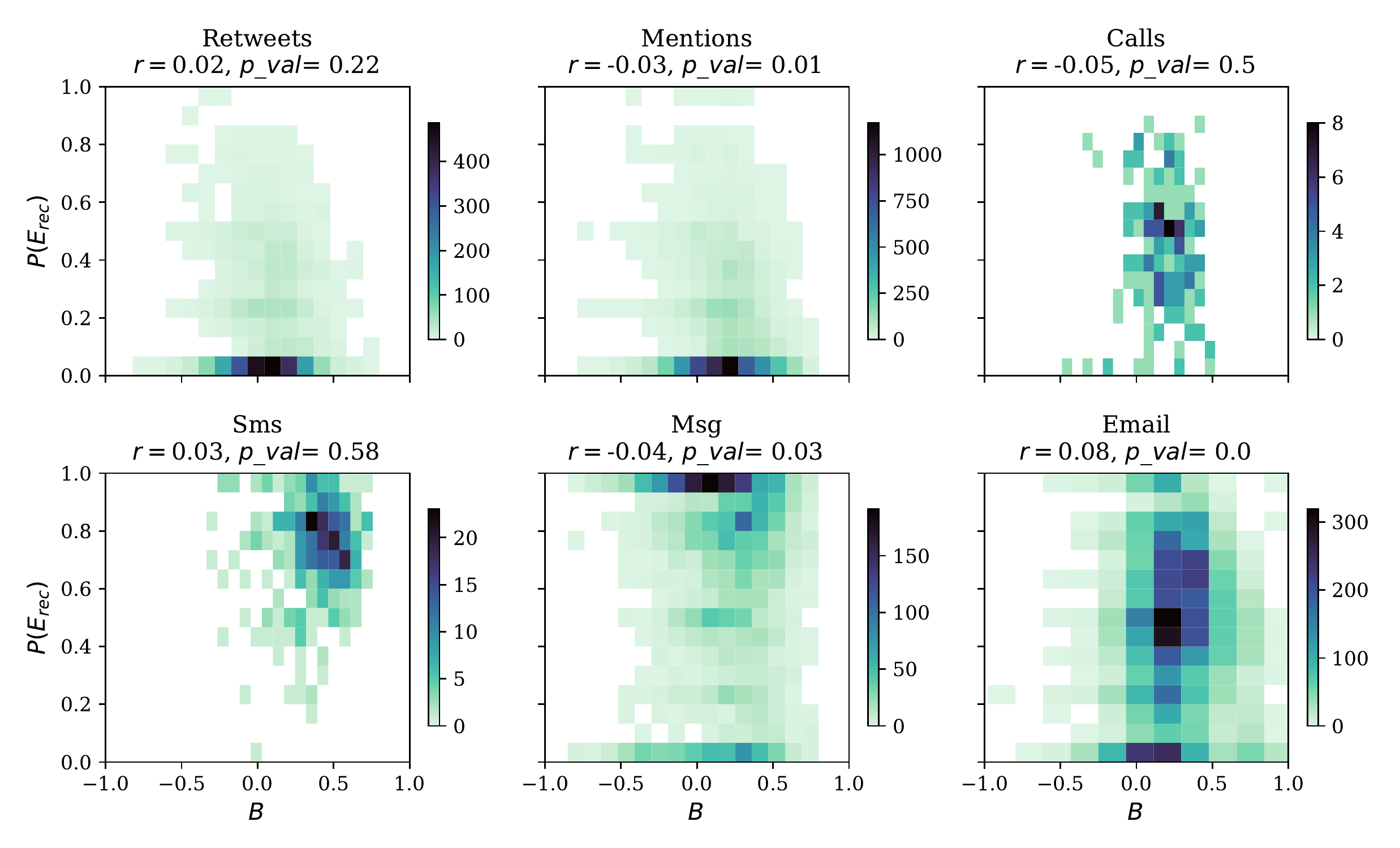}
\caption{Joint density of standard burstiness (here also denoted as $B$) and probability of reciprocation $\perec$ for all networks considered. We include the Pearson correlation coefficient ($r$) and the associated $p$-value ({\it p\_val}).}
\label{bvsR}
\end{figure}

\fref{bvsR} shows the joint density of standard burstiness (also denoted $B$, for simplicity) and the probability or reciprocation $\perec$ for each network considered. The Pearson correlation coefficient $r$ shows that there is no significant correlation between these two quantities. Thus, we need to investigate them separately.

\section{Reciprocation probability and balance}

The reciprocation probability $\perec$ between nodes $i$ and node $j$ has an upper bound related to balance $b$. Taking $n_{ij}$ and $n_{ji}$, respectively, as the number of events from $i$ to $j$ and from $j$ to $i$, we assume $n_{ij} < n_{ji}$ without loss of generality. The temporal configuration of events that maximizes $\perec$ is the one that maximizes the change of direction in interactions. In that case, the events from $i$ to $j$, the minority, are surrounded by events from $j$ to $i$, which creates two reciprocations by each event from $i$ to $j$. Then, the maximum probability of reciprocation is
\begin{equation}
p_{max}(E_{rec}) = \frac{2 \ min(n_{ij}, n_{ji})}{n_{ij}+ n_{ji}}.
\end{equation}
Since balance is defined as
\begin{equation}
b = \frac{max(n_{ij}, n_{ji})}{n_{ij}+ n_{ji}} = 1 - \frac{min(n_{ij}, n_{ji})}{n_{ij}+ n_{ji}},
\end{equation}
then the relation between the maximum probability of reciprocation and balance is
\begin{equation}
p_{max}(E_{rec}) = 2(1-b).
\end{equation}

\section{Effect of filtering in ADAM model}

%\san{Include details of fitting the activity, attractiveness from aggregated data, memory parameter from temporal dynamics of acquiring new neighbours. Also, ADA and ADAM follow different rules. But maybe we can say that we turn off memory for ADA?}
%\todo{Sandeep: Change $\ell$ to $l$ in axis label; change `Original' to `Data' in legend; change filter notation to $n_{events}=3$ in plot}

\begin{figure}[t]
\centering
\includegraphics[scale=0.6]{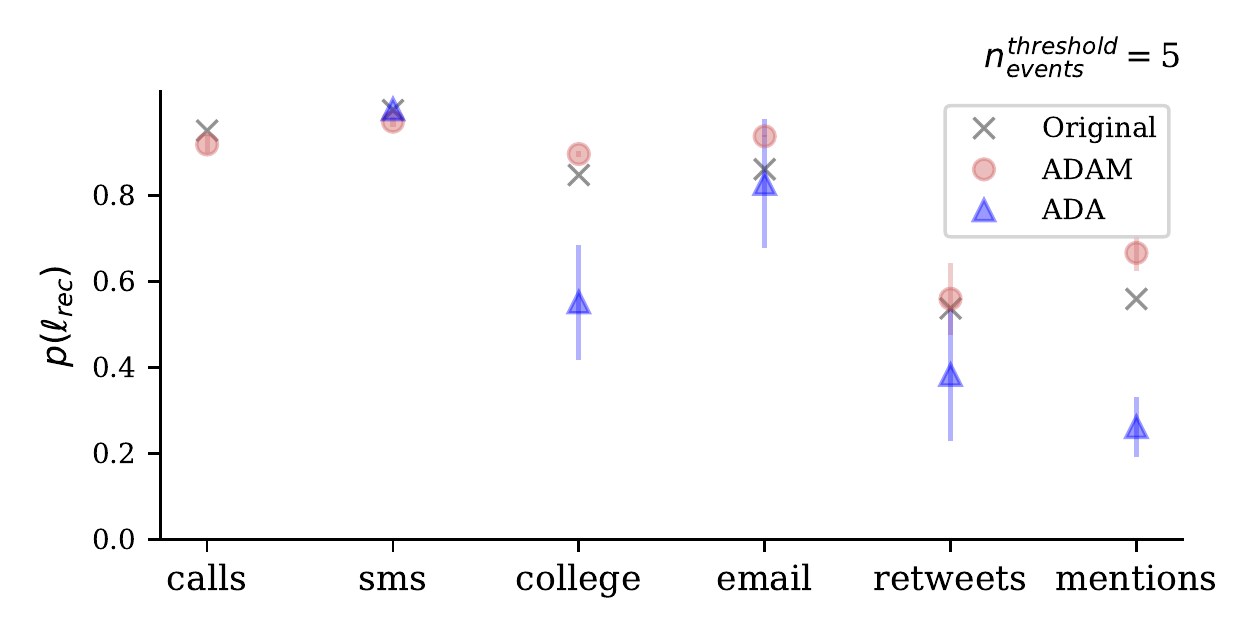}
\includegraphics[scale=0.6]{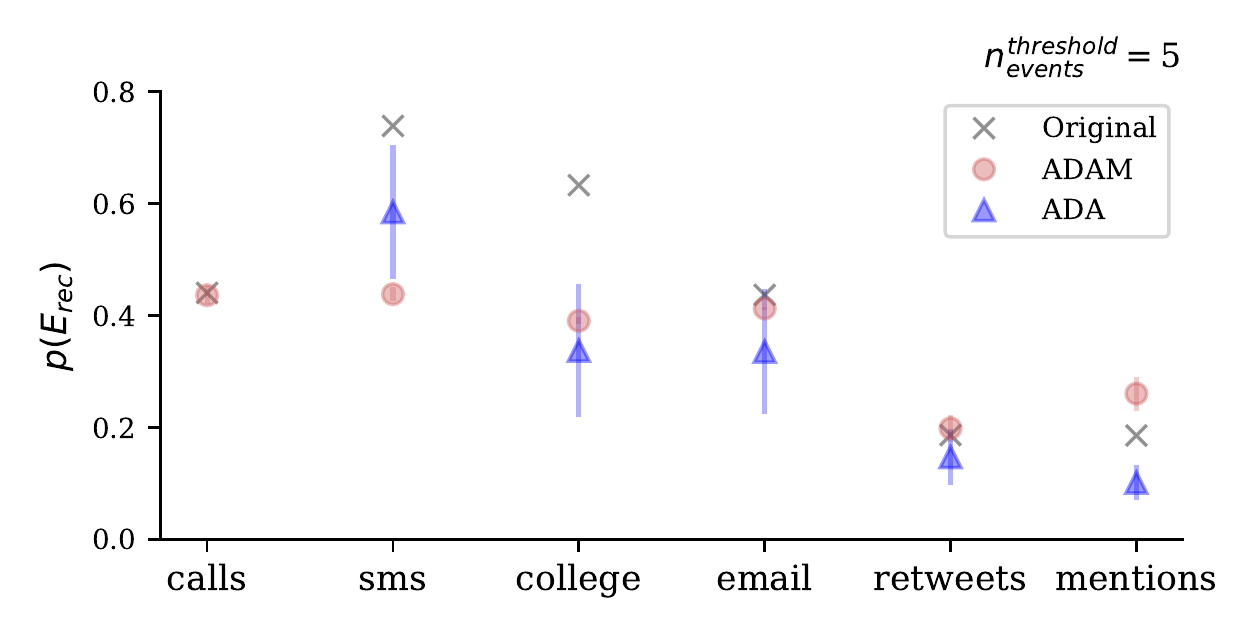}
\caption{Fraction of links having at least one reciprocation, $\plrec$ (left), and fraction of reciprocations $\perec$ (right), for several empirical communication channels (Original), as well as in synthetic temporal networks fitted by the ADAM and ADA models. The filter threshold is higher than in Fig. 4 in the main text (here denoted $n^{threshold}_{events}=5$).}
\label{model}
\end{figure}

Increasing the filter threshold to $n^{threshold}_{events}=5$ does not change qualitatively the results obtained in Fig. 4 of the main text (see \fref{model}).

\bibliographystyle{ieeetr}
\bibliography{refs}